\documentclass[conference]{IEEEtran}

\usepackage{graphicx}
\usepackage{amsmath,bbm,epsfig,amssymb,amsfonts, amstext, verbatim,amsopn,cite,subfigure,multirow,multicol,lipsum}
\usepackage{balance}
\usepackage{url}
\usepackage{amsfonts}
\usepackage{epsfig}
\usepackage{epstopdf}
\usepackage{setspace}
\usepackage{stmaryrd}
\usepackage{psfrag}	
\usepackage{multirow}
\usepackage{float}
\usepackage[process=auto]{pstool}
\usepackage{etoolbox}
\usepackage{algorithm}
\usepackage{algorithmic}
\usepackage{hyperref}

\usepackage{pifont}
\allowdisplaybreaks

\newtheorem{lem}{Lemma}

\newtheorem{prop}{Proposition}
\newtheorem{corol}{Corollary}

\newtoggle{ConfVersion}

\togglefalse{ConfVersion}

\iftoggle{ConfVersion}{%

\hoffset -2mm
\textwidth 18.5 cm
\textheight 24.95 cm

}{%

\hoffset -2mm
\textwidth 18.5 cm
\textheight 24.1 cm

}

\EndPreamble
\begin{document}

\title{Intelligent Reflecting Surfaces for Free~Space~Optical~Communications
\vspace{-0.35cm}}
\author{Marzieh Najafi$^{\dag}$ and Robert Schober$^{\dag}$\\  
$^{\dag}$Friedrich-Alexander University of Erlangen-Nuremberg, Germany
\vspace{-0.35cm}
}

\maketitle
\begin{abstract} 
In this paper, we investigate the use of intelligent reflecting surfaces (IRSs) (i.e., smart mirrors) to relax the line-of-sight requirement of free space optical (FSO) systems. We characterize the impact of the physical parameters of the IRS, such as its size, position, and orientation, on the quality of the end-to-end FSO channel. In addition, we develop a statistical channel model for the geometric and the misalignment losses which accounts for the random movements of the IRS, transmitter, and receiver due to building sway. This model can be used for performance analysis of IRS-based FSO systems. Our analytical results shows that depending on the angle between the beam direction and the IRS plane, building sway for the IRS has either a smaller or larger impact on the quality of the end-to-end FSO channel than building sway for the transmitter and receiver. Furthermore, our simulation results validate the accuracy of the developed channel model and offer insight for system design. 
\end{abstract}

\section{Introduction}
Intelligent reflecting surfaces (IRSs) have drawn considerable attention recently since they can be used to alter the radio frequency (RF) wireless channel for improved communication perfromance \cite{Rui_Zhang_RS,MetaSurf_Akyldiz,Vahid_RS,Alex_RS}. For example, IRSs have been used to extend the coverage of wireless communication systems to blind spots \cite{Rui_Zhang_RS,Alouini_RS} and to increase their security by improving the channel quality of the legitimate link and deteriorating the channel quality of the eavesdropper link \cite{Alex_RS}. Furthermore, IRSs are energy- and cost-efficient since they are composed of passive elements and can be installed on existing infrastructure, e.g., building walls. 

Optical wireless systems, e.g., free space optical (FSO) systems, are a promising candidate to meet the high data rate requirements of the next generation of wireless systems and beyond \cite{FSO_Survey_Murat,Steve_pointing_error,Alouini_Pointing,ICC_2018}. FSO systems offer the large bandwidth needed for applications such as wireless backhauling, while their transceivers are relatively cheap compared to their RF counterparts and easy to implement. However, the main requirement for establishing an FSO link is the existence of a line-of-sight (LOS) between the transceivers  \cite{FSO_Survey_Murat}. To relax this restrictive requirement, in this paper, we propose to use IRSs (smart mirrors) in FSO systems. Similar to RF-based IRSs, the IRSs in FSO systems can be installed on the walls of buildings. 
In RF systems, IRSs have to be equipped with a large number of passive phase shifters in order to create a narrow beam and to adaptively change the direction of the reflected beam to track mobile users \cite{Rui_Zhang_RS,Alouini_RS,Vahid_RS,Alex_RS}. In FSO systems, simple mirrors can be used to efficiently redirect the beam with negligible scattering \cite{Datasheet_Mirror}. Moreover, intelligent mirrors (i.e., optical IRSs) are able to control the direction of the reflected beam. This can be accomplished either by mechanically  rotating the IRS or by electronically changing the wavefront using advanced optical metasurfaces \cite{OpticMetaSurf2,OpticMetaSurf3}. In this paper, we consider the former case. 

Employing reflecting surfaces (RSs) (mirrors) in FSO systems has been widely considered in the literature \cite{brandl2013optical,Mirror_UAV_Exp,FSO_Mirror}. Mostly, RSs are used in the transceiver architecture in order to guide the optical beam in a desired direction \cite{brandl2013optical}. Another example of optical RSs is the passive retro-reflector which reflects the incoming laser beam back to its source and the reflected beam is modulated to carry data, see \cite{Mirror_UAV_Exp} for an experimental demonstration of  a retro-reflector. Furthermore, in \cite{FSO_Mirror}, the concept of using IRSs in FSO links was presented as a cost-effective solution for backhauling of cellular systems. However, the focus of \cite{FSO_Mirror} was on network planning and the impact of IRSs on the FSO channel model was not studied.


In this paper, we characterize the FSO channel between a transmitter (Tx), an IRS, and a receiver (Rx) as a function of the area, position, and orientation of the IRS. In particular, we derive the geometric and misalignment losses (GML) of the end-to-end link, i.e., the Tx-to-IRS-to-Rx link. Moreover, since, in addition to the Tx and Rx, the IRS is also affected by random movements due to building sway, we develop a statistical channel model which accounts for the impact of building sway for all three nodes. This model can be used to analyse the perfomance of IRS-based FSO systems. 
Our simulation results validate the proposed channel model and offer insight for system design.

\section{Preliminaries}

\subsection{System Model}

We consider an FSO communication system, where a Tx wishes to communicate with an Rx via an FSO link. We assume that there is no LOS between Tx and Rx. Hence, communication is enabled with the help of an IRS which has a LOS to both the Tx and the Rx. In other words, we assume that the Tx has an aperture directed towards the IRS; the IRS reflects the optical beam that it receives to the Rx; and the Rx collects the optical energy with a photo detector (PD). 

\subsection{Channel Model}

We assume an intensity modulation/direct detection (IM/DD) FSO system, where the PD responds to changes in the received optical signal power \cite{FSO_Survey_Murat}. Moreover, we assume that background noise is the dominant noise source at the PD and therefore the noise is independent from the signal \cite{FSO_Survey_Murat}.
The received signal at the Rx, denoted by $y_s$, is given by
\begin{IEEEeqnarray}{lll}\label{Eq:signal}
	y_s=hx_s+n,
\end{IEEEeqnarray}
where $x_s\in \mathbb{R}^+$ is the transmitted optical symbol (intensity),  $n\in \mathbb{R}$ is the zero-mean real-valued additive white Gaussian shot noise with variance $\sigma_n^2$  caused by ambient light at the Rx, and $h\in \mathbb{R}^+$ denotes the FSO channel gain. Moreover, we assume an average power constraint $\mathbbmss{E}\{x_s\}\leq{P}$. 

The FSO channel coefficient, $h$, is affected by several phenomena and can be modeled as \cite{Steve_pointing_error} 
\begin{IEEEeqnarray}{lll}\label{Eq:channel}
	h=\eta h_p h_a h_g,
\end{IEEEeqnarray}
where $\eta$ is the responsivity of the PD and $h_p$, $h_a$, and $h_g$ represent the atmospheric loss, atmospheric turbulence induced fading, and GML, respectively. In particular, the atmospheric loss, $h_p$, represents the power loss over a propagation path due to absorption and scattering of the light by particles in the atmosphere. The atmospheric turbulence, $h_a$, is induced by inhomogeneities in the temperature and the pressure of the atmosphere \cite{FSO_Survey_Murat}. The GML, $h_g$, is caused by the divergence of the optical beam along the propagation distance and the misalignment of the laser beam line\footnote{The beam line is the line that connets the laser source with the center of the beam footprint.} and the PD center due to building sway\cite{FSO_Survey_Murat,FSO_Vahid}. In this paper, our goal is to mathematically determine the impact of the IRS on the quality of the FSO channel.

\subsection{Problem Statement}
The impact of IRS on the end-to-end FSO channel is reflected in $h_p$ and $h_g$ which will be disscussed in the following:

\textit{i) Quality of reflection:} In addition to reflection, practical IRSs may also absorb or scatter some fraction of the beam power. Let $\zeta$ denote the reflection efficiency, i.e., the fraction of power reflected by the IRS. For practical IRSs, $\zeta$ usually assumes values in the range $[0.95,1]$  \cite{Datasheet_Mirror}. The absorption at the IRS can be regarded as a part of the atmospheric loss $h_p$.


\textit{ii) Relative position, orientation, and size of IRS:} The relative position and orientation of the IRS with respect to (w.r.t.) the laser beam determines the distribution of the \textit{reflected} optical power in space. The relative position and orientation of the PD w.r.t. the IRS determines the fraction of this power collected by the PD. Moreover, the size of the IRS determines which part of the PD is covered by the reflected beam. These parameters affect the \textit{mean} of the GML $h_g$. 

\textit{iii) Building Sway:}  The IRS is affected by the random movements of the building that it is installed on. This further increases the beam misalignment and affects the statistics of the GML $h_g$. In other words, the building sway of the buildings on which the Tx, Rx, and IRS are installed creates randomness in $h_g$.
 
Based on the above discussion, quantifying the impact of the IRS on the end-to-end FSO channel reduces to characterizing the corresponding GML $h_g$. To do so, we develop both a conditional model that accounts for the position, orientation, and size of the IRS and a statistical model that accounts for the random fluctuations of the IRS position due to building sway. As is customary for the analysis of optical systems \cite{VLC}, we first consider a two dimensional (2D) system model. The impact of the position, orientation, and building sway on $h_g$ can also be observed in a 2D system model. We generalize our model to a 3D system model in Section V based on the insights gained from analyzing the 2D system model.

\begin{figure}[t]
	\centering
	\scalebox{0.9}{
		\pstool[width=1\linewidth]{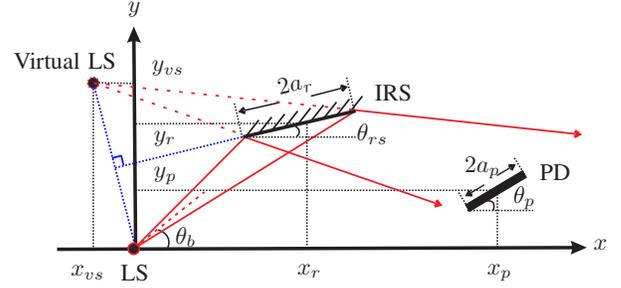}{
			\psfrag{A}[c][c][1]{$x$}
			\psfrag{C}[c][c][1]{$y$}
			\psfrag{xp}[c][c][1]{$x_p$}
			\psfrag{yp}[c][c][1]{$y_p$}
			\psfrag{ym}[c][c][1]{$y_r$}
			\psfrag{xm}[c][c][1]{$x_{r}$}
			\psfrag{ym1}[c][c][1]{$y_{vs}$}
			\psfrag{xm1}[c][c][1]{$x_{vs}$}
			\psfrag{M}[c][c][1]{IRS}
			\psfrag{P}[c][c][1]{PD}
			\psfrag{O}[c][c][1]{LS}
			\psfrag{v}[c][c][1]{Virtual LS}
            \psfrag{q}[c][c][1]{$\theta_{rs}$}
			\psfrag{z}[c][c][1]{$\theta_p$}
            \psfrag{t}[c][c][1]{$\theta_b$}
			\psfrag{L}[c][c][1]{$2a_p$}
			\psfrag{R}[c][c][1]{$2a_r$}
	}}
	\caption{2D schematic illustration of the considered IRS-based optical wireless system.}
	\label{Fig:Model}
\end{figure}

\section{Optical Power Collected by PD}

\subsection{Geometry of the Considered System}
We first define the position and orientation of the laser source (LS), the IRS, and the PD in the considered 2D system model, which is schematically illustrated in Fig.~\ref{Fig:Model}. 
Without loss of generality, we assume that the LS is located in the origin of the coordinate system, i.e., $\mathbf{p}_{\mathrm{ls}}=(0,0)$. The center of the IRS and the PD are located at $\mathbf{c}_{\mathrm{rs}}=(x_r,y_r)$ and $\mathbf{c}_{\mathrm{pd}}=(x_p,y_p)$, respectively. The lengths of the IRS and the PD are denoted by $2a_r$ and $2a_p$, respectively. The direction of the laser beam is determined by the angle between the laser beam line and the $x$ axis denoted by $\theta_b$. The IRS and the PD have angles $\theta_{rs}$ and $\theta_p$ w.r.t. the $x$ axis, repectively.  For convenience, the LS can be mirrored at the line defined by the IRS, cf. Fig.~\ref{Fig:Model}, and the resulting virtual LS can be used in the subsequent analysis \cite{Phys}. The virtual laser beam has an angle of $2\theta_{rs}-\theta_b$ with the $x$ axis and the position of the virtual LS is given by 
\begin{IEEEeqnarray}{lll}
\mathbf{p}_{\mathrm{vs}}=(x_{vs}, y_{vs})=&\Big(\big(1-\cos(2\theta_{rs})\big)x_r-\sin(2\theta_{rs})y_r,\nonumber\\ &\big(1+\cos(2\theta_{rs})\big)y_r-\sin(2\theta_{rs})x_r\Big).
\end{IEEEeqnarray}

\subsection{Spatial Distribution of the Reflected Power Density}

Next, we derive the power density of the reflected beam across space. We assume a Gaussian beam which dictates that the power density distribution across any line perpendicular to the direction of the wave propagation follows a Gaussian profile \cite{FSO_Survey_Murat,Steve_pointing_error}. Let us consider a line that is perpendicular to the beam direction and the distance between the center of the beam footprint on the line and the LS is denoted by $d$. Then, the power density for any point on this perpendicular line with distance $r$ from the center of the beam footprint is given by \cite{Steve_pointing_error}
\begin{IEEEeqnarray}{lll} \label{Eq:PowerOrthogonal}
I^{\mathrm{orth}}(r;d) = \frac{2}{\pi w^2(d)}\exp\left(-\frac{2r^2}{w^2(d)}\right),
\end{IEEEeqnarray}
where $w(d)$ is the beam width at distance $d$ and is given by
\begin{IEEEeqnarray}{lll} \label{Eq:BeamWidth}
w(d)= w_0\sqrt{1+\left(1+\frac{2w_0^2}{\rho^2(d)}\right)\left(\frac{\lambda d}{\pi w_0^2}\right)^2}.
\end{IEEEeqnarray}
Here, $w_0$ denotes the beam waist radius, $\rho(d)=(0.55C_n^2k^2d)^{-3/5}$ is referred to as the coherence length, $k=\frac{2\pi}{\lambda}$ is the wave number, $\lambda$ denotes the optical wavelength, and $C_n^2\approx C_0\exp\left(-\frac{h}{100}\right)$ is the index of refraction structure parameter, where $C_0=1.7\times10^{-14}$~m$^{\frac{2}{3}}$ is the nominal value of the refractive index at the ground and $h$ is the operating height of the FSO transceivers \cite{FSO_Survey_Murat}. The following lemma provides the power density of the \textit{beam reflected} by the IRS.

\begin{lem}\label{Lem:Truncated Gauss}
Assuming a transmitted Gaussian beam, the power density of the reflected beam on a perpendicular line w.r.t. the beam direction at point $\tilde{\mathbf{p}}=(\tilde{x},\tilde{y})$ is given by
\begin{IEEEeqnarray}{lll} \label{Eq:Truncated Gauss}
I^{\mathrm{orth}}_{\mathrm{rfl}}(\tilde{\mathbf{p}})=\begin{cases}
\frac{2}{\pi w^2(d)}\exp\left(-\frac{2r^2}{w^2(d)}\right), &\tilde{\mathbf{p}}\in\mathcal{R}\\
0, &\mathrm{otherwise},
\end{cases}
\end{IEEEeqnarray}
where $r=\|\tilde{\mathbf{p}}-\mathbf{p}_{\mathrm{c}}\|$ and $d=\|\mathbf{p}_{\mathrm{vs}}-\mathbf{p}_{\mathrm{c}}\|$ are the distances between the beam footprint center and point $\tilde{\mathbf{p}}$ and the LS, respectively. Region $\mathcal{R}$ is defined as $\mathcal{R}=\{(x,y)|s_1(x-x_{vs})+y_{vs}\leq y\leq s_2(x-x_{vs})+y_{vs}\}$, where $s_1=\frac{y_r-\sin(\theta_{rs})a_r-y_{vs}}{x_r-\cos(\theta_{rs})a_r-x_{vs}}$ and 
$s_2=\frac{y_r+\sin(\theta_{rs})a_r-y_{vs}}{x_r+\cos(\theta_{rs})a_r-x_{vs}}$ and $\mathbf{p}_{\mathrm{c}}$ is given by
\begin{IEEEeqnarray}{lll} \label{Eq:xcyc}
\mathbf{p}_{\mathrm{c}}^{\mathsf{T}}
=&
\begin{bmatrix}
\tan(\theta_b-2\theta_{rs}) &1\\
-\cot(\theta_b-2\theta_{rs})& 1
\end{bmatrix}^{-1}
\begin{bmatrix}
\tan(\theta_b-2\theta_{rs})x_{vs}+y_{vs}\\
-\cot(\theta_b-2\theta_{rs})\tilde{x}+\tilde{y}
\end{bmatrix}.\nonumber\\
\end{IEEEeqnarray}
\end{lem}

\begin{IEEEproof}
\iftoggle{ConfVersion}{%
Due to space constraints, the proof is given in \cite[Appendix~A]{Globecom2019_arXiv}, which is an extended version of this paper\footnote{Globecom 2019 allows only 6-page manuscripts for review but the final paper can have 7 pages. If the paper is accepted, we will submit the extended version \cite{Globecom2019_arXiv} including the proofs as the final paper.}.
}{%
Please refer to Appendix~\ref{App:Lem_Truncated Gauss}.
}
\end{IEEEproof}

Lemma~\ref{Lem:Truncated Gauss} provides several insights regarding the impact of the IRS on the reflected power distribution. In particular, the reflected beam is a \textit{truncated} Gaussian beam which originates from the virtual LS and is confined to area $\mathcal{R}$. Moreover, the size of $\mathcal{R}$ depends on the size of the IRS as well as on its relative orientation w.r.t. the laser beam. Furthermore, for a given $d$, $I^{\mathrm{orth}}_{\mathrm{rfl}}(\tilde{\mathbf{p}})$ attains its maximum, i.e., $\frac{2}{\pi w^2(d)}$, at $r=0$, i.e., at the center of its footprint $\mathbf{p}_{\mathrm{c}}$, cf. \eqref{Eq:Truncated Gauss}. Note that $\mathbf{p}_{\mathrm{c}}$ depends on $\theta_{rs}$ and the value of point $\tilde{\mathbf{p}}$. Therefore, for $\tilde{\mathbf{p}}$ on the PD, for an efficient design, we should choose $\theta_{rs}$ such that $\mathbf{p}_{\mathrm{c}}$ lies in the center of the PD $\mathbf{c}_{\mathrm{pd}}=(x_p,y_p)$ and distance $d$ is the end-to-end distance between the LS and the PD, $d=d_{sr}+d_{rp}\triangleq d_{e2e}$, where $d_{sr}$ and $d_{rp}$ denote the distances between the LS to the IRS and the IRS to the PD, respectively. This leads to the optimal $\theta_{rs}^*$ which is found as the unique solution of the following equation
\begin{IEEEeqnarray}{lll} \label{Eq:OptimalAngle}
\tan(2\theta_{rs}^*-\theta_b)
=\frac{(1+\cos(2\theta_{rs}^*))y_r-\sin(2\theta_{rs}^*)x_r-y_p}{(1-\cos(2\theta_{rs}^*))x_r-\sin(2\theta_{rs}^*)y_r-x_p}.\quad
\end{IEEEeqnarray}

\subsection{Conditional GML Model}

In order to compute the GML, we have to integrate the reflected power density over the PD, i.e.,  
\begin{IEEEeqnarray}{lll} \label{Eq:Integral}
	h_g= \int_{\tilde{\mathbf{p}}\in\mathcal{P}} \sin(\psi)I^{\mathrm{orth}}_{\mathrm{rfl}}(\tilde{\mathbf{p}}) \mathrm{d}\tilde{\mathbf{p}},
\end{IEEEeqnarray}
where $\psi=\theta_b+\theta_p-2\theta_{rs}$ is the angle between the PD and the beam line and $\mathcal{P}$ is the set of points on the PD, i.e., 
\begin{IEEEeqnarray}{lll}\mathcal{P}=&\Big\{(x,y)|y=\tan(\theta_p)(x-x_p)+y_p, x\in\big[x_p-\cos(\theta_{p})a_p,\nonumber\\
&x_p+\cos(\theta_{p})a_p\big], y\in\big[y_p-\sin(\theta_{p})a_p,y_p+\sin(\theta_{p})a_p\big]\Big\}.\nonumber\\
\end{IEEEeqnarray}
The term $\sin(\psi)\in[0,1]$ in \eqref{Eq:Integral} accounts for the non-orthogonality of the PD. Let $L_c$ denote the distance between the center of the PD and the beam line, i.e., $L_c=\|\mathbf{c}_{\mathrm{pd}}-\mathbf{p}_{\mathrm{c}}\|$ for $d=d_{e2e}$. The following proposition provides a closed-form expression for the GML $h_g$. For future reference, $\mathbf{a}\leq \mathbf{b}$ indicates that all elements of $\mathbf{a}$ are smaller than the corresponding elements in $\mathbf{b}$.

\begin{prop}\label{Prop:TotalFrac}
Under the mild condition $a_p,L_c\ll d_{e2e}$, the total fraction of power that is captured by the PD  is given by
\begin{IEEEeqnarray}{lll} \label{Eq:Pt}
h_g=\frac{1}{\sqrt{2\pi}w(d_{e2e})}\times\\
\begin{cases}\displaystyle
\mathrm{erf}\bigg(\frac{\sqrt{2}\sin(\psi)\rho_1}{w(d_{e2e})}\bigg)+\mathrm{erf}\bigg(\frac{\sqrt{2}\sin(\psi)\rho_2}{w(d_{e2e})}\bigg),\hspace{-2mm}&\mathrm{if}\,\,\rho_{12}=2a_p\\
\displaystyle
\bigg|\mathrm{erf}\bigg(\frac{\sqrt{2}\sin(\psi)\rho_1}{w(d_{e2e})}\bigg)-\mathrm{erf}\bigg(\frac{\sqrt{2}\sin(\psi)\rho_2}{w(d_{e2e})}\bigg)\bigg|,\hspace{-2mm}&\mathrm{otherwise,}
\end{cases}\nonumber
\end{IEEEeqnarray}
where $\rho_1=\|\mathbf{p}_0-\hat{\mathbf{p}}_1\|$, $\rho_2=\|\mathbf{p}_0-\hat{\mathbf{p}}_2\|$, and $\rho_{12}=\|\mathbf{p}_0-\mathbf{p}_1\|+\|\mathbf{p}_0-\mathbf{p}_2\|$. Moreover, $\mathbf{p}_0$, $\mathbf{p}_1$,  $\hat{\mathbf{p}}_1$, $\mathbf{p}_2$, and $\hat{\mathbf{p}}_2$ are given in \eqref{Eq:r_1} at the top of the next page. 
\begin{figure*}
\begin{IEEEeqnarray}{ccc} \label{Eq:r_1}
\mathbf{p}_0^{\mathsf{T}}
=
\begin{bmatrix}
\tan(\theta_b-2\theta_{rs}) &1\\
-\tan(\theta_p)& 1
\end{bmatrix}^{-1}
\begin{bmatrix}
\tan(\theta_b-2\theta_{rs})x_{vs}+y_{vs}\\
-\tan(\theta_p)x_p+y_p
\end{bmatrix},\quad
\mathbf{p}_1^{\mathsf{T}}
=
\begin{bmatrix}
x_p+a_p\cos(\theta_p)\\
y_p+a_p\sin(\theta_p)
\end{bmatrix},\quad
\widetilde{\mathbf{p}}_1^{\mathsf{T}}
=
\begin{bmatrix}
-s_2 &1\\
-\tan(\theta_p)& 1
\end{bmatrix}^{-1}\times\nonumber\\
\begin{bmatrix}
-s_2x_{vs}+y_{vs}\\
-\tan(\theta_p)x_p+y_p
\end{bmatrix},\quad
\mathbf{p}_2^{\mathsf{T}}
=
\begin{bmatrix}
x_p-a_p\cos(\theta_p)\\
y_p-a_p\sin(\theta_p)
\end{bmatrix},\quad
\widetilde{\mathbf{p}}_2^{\mathsf{T}}
=
\begin{bmatrix}
-s_1 &1\\
-\tan(\theta_p)& 1
\end{bmatrix}^{-1}
\begin{bmatrix}
-s_1x_{vs}+y_{vs}\\
-\tan(\theta_p)x_p+y_p
\end{bmatrix},\nonumber\\
\hat{\mathbf{p}}_1
=
\begin{cases}
\widetilde{\mathbf{p}}_2,  \quad &  \mathbf{p}_1<\widetilde{\mathbf{p}}_2\\
\mathbf{p}_1,  &  \widetilde{\mathbf{p}}_2\leq\mathbf{p}_1\leq \widetilde{\mathbf{p}}_1\\
\widetilde{\mathbf{p}}_1,  &  \mathbf{p}_1>\widetilde{\mathbf{p}}_1
\end{cases},\quad
\hat{\mathbf{p}}_2
=
\begin{cases}
\widetilde{\mathbf{p}}_2, \quad  &  \mathbf{p}_2<\widetilde{\mathbf{p}}_2\\
\mathbf{p}_2,   &  \widetilde{\mathbf{p}}_2\leq\mathbf{p}_2\leq \widetilde{\mathbf{p}}_1\\
\widetilde{\mathbf{p}}_1,  &  \mathbf{p}_2>\widetilde{\mathbf{p}}_1.
\end{cases}
\end{IEEEeqnarray}
\noindent\rule{1\linewidth}{0.4pt}
\end{figure*}
\end{prop}
\begin{IEEEproof}
\iftoggle{ConfVersion}{%
Due to space constraints, the proof is given in \cite[Appendix~B]{Globecom2019_arXiv}, which is an extended version of this paper.
}{%
Please refer to Appendix~\ref{App:Prop_TotalFrac}.
}
\end{IEEEproof}

Note that the conditions under which \eqref{Eq:Pt} in Proposition~\ref{Prop:TotalFrac} holds are met in practice since 1) the physical size of the PD is much smaller than the transmission distance, i.e., $a_p\ll d_{e2e}$ holds, and 2)  $L_c$ corresponds to the beam misalignment and for a properly designed system, the misalignment is much smaller than the end-to-end transmission distance, i.e.,  $L_c\ll d_{e2e}$ holds. The impact of the size of the IRS is reflected in the values of $\rho_{1}$ and $\rho_{2}$. In fact, if the IRS is sufficiently large such that the PD is located in region $\mathcal{R}$ defined in Lemma~\ref{Lem:Truncated Gauss}, we obtain $\rho_1=\|\mathbf{p}_0-\mathbf{p}_1\|$, $\rho_2=\|\mathbf{p}_0-\mathbf{p}_2\|$.

\begin{corol}\label{Corol:Perp}
For the special case where $\hat{\mathbf{p}}_i=\mathbf{p}_i, i=1,2$, i.e., the IRS is sufficiently large, and the reflected beam strikes the center of the PD and its direction is perpendicular to the PD, the total fraction of power that is captured by the PD is obtained as
\begin{IEEEeqnarray}{lll} \label{Eq:PtCorol}
h_g=\frac{\sqrt{2}}{\sqrt{\pi}w(d_{e2e})}\mathrm{erf}\Big(\frac{\sqrt{2}a_p}{w(d_{e2e})}\Big).
\end{IEEEeqnarray}
\end{corol}
\begin{IEEEproof}
Eq. \eqref{Eq:PtCorol} is obtained by substituting $\psi=\frac{\pi}{2}$ and $\rho_1=\rho_2=\|\mathbf{p}_0-\mathbf{p}_1\|=\|\mathbf{p}_0-\mathbf{p}_1\|=a_p$ into \eqref{Eq:Pt}. This completes the proof.  
\end{IEEEproof}

For a given end-to-end distance $d_{e2e}$ and a given PD area $a_p$, the maximum fraction of power collected by the PD is given by \eqref{Eq:PtCorol}. To attain this maximum, three conditions have to hold, namely the IRS is sufficiently large, the misalignment is zero,  i.e., $\theta_{rs}=\theta_{rs}^*$, cf. \eqref{Eq:OptimalAngle}, and the PD is orthogonal to the beam line, i.e., $\theta_p=\frac{\pi}{2}+2\theta_{rs}-\theta_b$.

\section{Statistical Model - 2D System}
In this section, we study the effect that building sway has on the quality of the considered FSO channel.

\subsection{Building Sway Model}
We assume that the positions of the LS, IRS, and PD fluctuate because of building sway in both the $x$ and $y$ directions. In the following, we show that for the LS, IRS, and PD only the fluctuations in a certain direction have a considerably impact on the FSO channel, respectively. This observation substantially simplifies the derivation of a statistical channel model. 

\textbf{LS:} The fluctuations of the position of the LS can be projected in the beam direction and the direction orthogonal to it. Let $\epsilon_s^b$ and $\epsilon_s^o$ denote the fluctuations of the LS position for the former and latter cases, respectively. Hereby, since the fluctuations of the LS in the beam direction are much smaller than the distance between the LS and the IRS, the impact of $\epsilon_s^b$ on $h_g$ can be safely neglected.

\textbf{IRS:} The fluctuations of the position of the IRS can be projected in the direction along the IRS line and the orthogonal direction denoted by $\epsilon_r^{r}$ and $\epsilon_r^{o}$, respectively. Assuming that the beam line is aligned to pass through the IRS (not necessarily its center) and that the size of the IRS is large, the impact of $\epsilon_r^{r}$ on $h_g$ is negligible. Nevertheless, $\epsilon_r^{o}$ may considerably change the position of the beam footprint center at the PD.  

\textbf{PD:} Similar to the LS, let $\epsilon_p^b$ and $\epsilon_p^o$ denote the fluctuations of the position of the PD in the direction of the reflected beam and   perpendicular to it, respectively. Since the distance between the IRS and the PD is much larger than the fluctuations in the reflected beam direction, we can safely neglect the impact of $\epsilon_p^b$ on $h_g$. 

Let $u$ denote misalignment between the center of the beam footprint and the center of the PD. $u$ is given in the following lemma. 
\begin{lem}\label{Lem:Misalignment}
The misalignment $u$  as a function of $(\epsilon_s^o,\epsilon_r^o,\epsilon_p^o)$ is obtained as
\begin{IEEEeqnarray}{lll} \label{Eq:Misalignment}
	u=\frac{1}{\sin(\psi)}(\epsilon_s^o+2\cos(\theta_b-\theta_{rs})\epsilon_r^o+\epsilon_p^o).
\end{IEEEeqnarray}
\end{lem}
\begin{IEEEproof}
In \eqref{Eq:Misalignment}, the term $\epsilon_s^o+2\cos(\theta_b-\theta_{rs})\epsilon_r^o+\epsilon_p^o$ captures the misalignment on a plane perpendicular to the direction of the reflected beam and the term $\frac{1}{\sin(\psi)}$ accounts for the non-orthogonality of the PD. Moreover, the fluctuations of the LS and PD are projected onto the perpendicular misalignment without any change, whereas the projection of the fluctuations of the IRS onto the perpendicular misalignment depends on angle $\theta_b-\theta_{rs}$ as given in \eqref{Eq:Misalignment}.  This completes the proof.  
\end{IEEEproof}

Note that $(\epsilon_s^o,\epsilon_r^o,\epsilon_p^o)$ are random variables (RVs). A widely-accepted model for building sway assumes independent zero-mean Gaussian fluctuations \cite{Steve_pointing_error,ICC_2018}, i.e., $\epsilon_s^o\sim\mathcal{N}(0,\sigma^2_s)$, 
 $\epsilon_r^o\sim\mathcal{N}(0,\sigma^2_r)$, and $\epsilon_p^o\sim\mathcal{N}(0,\sigma^2_p)$, where $\sigma_i^2$ denotes the variance of $\epsilon_i^o,\,\,i\in\{s,r,p\}$. Therefore, the misalignment also follows a zero-mean Gaussian distribution, i.e., $u\sim\mathcal{N}\big(0,\sigma^2\big)$ with variance $\sigma^2=\frac{1}{\sin^2(\psi)}(\sigma_s^2+4\cos^2(\theta_b-\theta_{rs})\sigma_r^2+\sigma_p^2)$.

\subsection{PDF of Power Collected by the PD}
In order to derive the statistical channel model for the GML $h_g$, first the power collected by the PD has to be derived as a function of $u$. To do so, we can use the exact expressions in \eqref{Eq:Pt} and replace $(\rho_1,\rho_2)$ with $(|u-a_p|,u+a_p)$, assuming that the IRS is sufficiently large such that the PD is located in region $\mathcal{R}$ defined in Lemma~\ref{Lem:Truncated Gauss}. However, the resulting expressions are rather complicated and do not provide useful insights. Thus, to get some insights, we approximate $h_g$ as a function of $u$ as follows
\begin{IEEEeqnarray}{lll} \label{Eq:A0Approx}
h_g\approx A_0 \exp\Big(\frac{-2u^2}{tw^2(d_{e2e})}\Big),
\end{IEEEeqnarray}
where $A_0=\frac{\sqrt{2}}{\sqrt{\pi}w(d_{e2e})}\mathrm{erf}\left(\nu\right)$, $t=\frac{\sqrt{\pi}\mathrm{erf}\left(\nu\right)}{2\nu\exp(-\nu^2)\sin^2(\psi)}$, and $\nu=\frac{\sqrt{2}\sin(\psi)a_p}{w(d_{e2e})}$.
The derivation of \eqref{Eq:A0Approx} is provided in \iftoggle{ConfVersion}{\cite[Appendix~C]{Globecom2019_arXiv}}{Appendix~\ref{App:A0Approx}}. We verify the accuracy of \eqref{Eq:A0Approx} in Section~VI. Using this approximation, the PDF of $h_g$ is given in the following proposition.

\begin{prop}\label{Prop:2}
Based on \eqref{Eq:Misalignment} and \eqref{Eq:A0Approx} and assuming Gaussian fluctuations,  $h_g$ follows a distribution with the following PDF
\begin{IEEEeqnarray}{lll} \label{Eq:PDFs}
f_{h_g}(h_g)=\frac{\varpi}{2A_0\sqrt{\pi}}\left[\ln\Big(\frac{A_0}{h_g}\Big)\right]^{-\frac{1}{2}}&\left(\frac{h_g}{A_0}\right)^{\varpi-1},\nonumber\\
& 0\leq h_g\leq A_0.\qquad
\end{IEEEeqnarray}
where $\varpi=\frac{tw^2(d_{e2e})}{4\sigma^2}$.
\end{prop}
\begin{IEEEproof}
Eq. \eqref{Eq:PDFs} can be obtained by exploiting the relation between the PDF of $u$ and $h_g$ in \eqref{Eq:A0Approx} and the fact that $u$ follows a zero-mean Gaussian distribution.
\end{IEEEproof}

Proposition \ref{Prop:2} reveals the impact of system parameters such as $d_{e2e}$ and $\sigma^2$ on the PDF of the GML.

\section{Extension to 3D System Model}
For the 2D system model, we needed two position variables and one angular variable to characterize the positions and orientations of the LS, IRS, and PD, respectively, i.e., in total 9 parameters. In contrast, for a 3D system model, we require three position variables and two angular variables to characterize the positions and orientations of the nodes, i.e., in total 15 parameters. This severely complicates the analysis of the 3D system. To cope with this issue, we exploit the insights gained from analyzing the 2D system and characterize the 3D system only w.r.t. those parameters that affect the GML $h_g$.  From Sections~II-IV, we offer the following observations:

\begin{itemize}
\item Lemma~\ref{Lem:Truncated Gauss} reveals that in 2D systems, the impact of the IRS can be modeled via a virtual LS where the reflected beam follows a truncated Gaussian profile. The position of the virtual LS depends on the relative position and orientation of the IRS w.r.t. the beam line. Nevertheless, the distance between the virtual LS and the PD is the sum of the distances between the actual LS to the IRS and the IRS to the PD, i.e., $d_{e2e}$. Moreover, the truncation can be ignored if the IRS is sufficiently large such that the PD is completely inside region $\mathcal{R}$ defined in Lemma~\ref{Lem:Truncated Gauss}.
\item The conditional model in \eqref{Eq:A0Approx} reveals that the overall impact of the position and orientation parameters of the IRS and the PD on the GML $h_g$ manifests itself in three variables, namely misalignment $u$, end-to-end distance $d_{e2e}$, and angle $\psi$. Due to building sway, the misalignment $u$ is an RV; however, by a proper system design, i.e., by choosing $\theta_{rs}=\theta^*_{rs}$ according to \eqref{Eq:OptimalAngle}, one can make the average misalignment $u$ vanish, i.e., $\mathbbmss{E}\{u\}=0$. 
\end{itemize}

In the following, we exploit the two above observations for analyzing a 3D system. Let $\psi_p$ denote the angle between the reflected beam and the PD \textit{plane}. Assuming a circular PD of radius $a_p$, the following approximate expression was recently obtained in \cite{ICC_2018} for the GML of a 3D system
\begin{IEEEeqnarray}{lll} \label{Eq:hg_3D}
h_g(u) \approx A_0 \exp\left(-\frac{2\|\mathbf{u}\|^2}{tw^2(d_{e2e})}\right),
\end{IEEEeqnarray}
where $t=\sqrt{t_1t_2}$, $t_1=\frac{\sqrt{\pi}\mathrm{erf}(\nu_1)}{2\nu_1\exp(-\nu_1^2)}$ , $t_2=\frac{\sqrt{\pi}\mathrm{erf}(\nu_2)}{2\nu_2\exp(-\nu_2^2)\sin^2(\psi_p)}$,  $\nu_{1}=\frac{a_p}{w(d_{e2e})}\sqrt{\frac{\pi}{2}}$, and  $\nu_{2}=\nu_{1}|\sin(\psi_p)|$. Moreover, $\mathbf{u}$ denotes the vector of misalignment on the PD plane, and $A_0$ denotes the maximum fraction of optical power captured by the PD at $\|\mathbf{u}\|=0$ and is given by $A_0=\mathrm{erf}(\nu_{1})\mathrm{erf}(\nu_{2})$.  Note that the exact expression for $h_g$ can be obtained in a similar manner as that obtained in Proposition~\ref{Prop:TotalFrac} for 2D systems but is much more involved. In the following, we derive a statistical model based on \eqref{Eq:hg_3D} incorporating the impact of the IRS.

Similar to the statistical analysis for 2D systems given in Section~IV,  we assume Gaussian fluctuations due to building sway for the LS, IRS, and PD as described in the following. 

\textbf{LS:} In general, the fluctuations of the position of a point in a 3D system can be modeled by three variables in three orthogonal directions. For the LS, fluctuations along the direction of the beam have negligible impact on $h_g$; hence, we need only two  variables in two orthogonal directions on the plane perpendicular to the beam direction, denoted by $\epsilon_s^{o_1},\epsilon_s^{o_2}\sim\mathcal{N}(0,\sigma_s^2)$.

\textbf{IRS:} Since we assume a sufficiently large IRS, the fluctuations of the IRS along its plane can be neglected. Therefore, we need to consider only the fluctuations orthogonal to the IRS plane, denoted by $\epsilon_r^{o}\sim\mathcal{N}(0,\sigma_r^2)$. 

\textbf{PD:} Similar to the LS, the fluctuations along the reflected beam direction can be neglected. Hence, we need two variables in two orthogonal directions to describe the fluctuations in the plane perpendicular to the reflected beam, denoted by $\epsilon_p^{o_1},\epsilon_p^{o_2}\sim\mathcal{N}(0,\sigma_p^2)$.

It is interesting to note that the GML is affected by the IRS only via variable $\epsilon_r^{o}$. This implies that variations of $\epsilon_r^{o}$ lead to variations of $\mathbf{u}$ along only one dimension. Without loss of generality and to simplify our notation, we choose the basis for variables $(\epsilon_s^{o_1},\epsilon_s^{o_2})$ and $(\epsilon_p^{o_1},\epsilon_p^{o_2})$ such that the variations of $\mathbf{u}$ due to $\epsilon_s^{o}$ and $\epsilon_p^{o}$ are in the same direction as those due to $\epsilon_r^{o}$. Based on this convention, the following lemma presents the misalignment vector $\mathbf{u}$. 

\begin{lem}\label{Lem:Misalignment3D} 
	The misalignment vector $\mathbf{u}$  as a function of $(\epsilon_s^{o_1},\epsilon_s^{o_2})$, $\epsilon_r^{o}$, and $(\epsilon_p^{o_1},\epsilon_p^{o_2})$ is obtained as
	\begin{IEEEeqnarray}{lll} \label{Eq:Misalignment3D}
		\mathbf{u}=\frac{1}{\sin(\psi_p)}\left(\epsilon_s^{o_1}+2\cos(\psi_{r})\epsilon_r^o+\epsilon_p^{o_1}, \epsilon_s^{o_2}+\epsilon_p^{o_2}\right),
	\end{IEEEeqnarray}
	where $\psi_{r}$ is the angle between the laser beam and the IRS plane. 
\end{lem}
\begin{IEEEproof}
	The proof is similar to that given for Lemma~\ref{Lem:Misalignment} for 2D systems. The convention for the definition of the bases for $(\epsilon_s^{o_1},\epsilon_s^{o_2})$ and $(\epsilon_p^{o_1},\epsilon_p^{o_2})$ facilitates the derivation of $\mathbf{u}$ since $\epsilon_r^o$ affects only one of the dimensions of $\mathbf{u}$.
\end{IEEEproof}
  
 Assuming $\mathbf{u}=(u_1,u_2)$, $u_1$ and $u_2$ follow Gaussian distributions with zero mean and variances $\sigma_{u_1}^2=\frac{1}{\sin^2(\psi_p)}(\sigma_s^2+4\cos^2(\psi_r)\sigma_r^2+\sigma_p^2)$ and $\sigma_{u_2}^2=\frac{1}{\sin^2(\psi_p)}(\sigma_s^2+\sigma_p^2)$, respectively. Therefore, $\|\mathbf{u}\|$ follows a Hoyt distribution which is given by \cite{Alouini_Pointing}
\begin{IEEEeqnarray}{lll} \label{Eq:PDFu3D}
f_{\|\mathbf{u}\|}(u)=\frac{1+q^2}{q\Omega}u\exp\left(-\frac{(1+q^2)^2}{4q^2\Omega}u^2\right)I_0\left(\frac{1-q^4}{4q^2\Omega}u^2\right),\nonumber\\
\end{IEEEeqnarray}
where $q=\frac{\sigma_{u_2}}{\sigma_{u_1}}$, $\Omega=\sigma_{u_1}^2+\sigma_{u_2}^2$, and $I_0(\cdot)$ is the zero-order modified Bessel function of the first kind.
For the special case where $\sigma_r^2=0$, $\|\mathbf{u}\|$ is Rayleigh distributed, similar to the pointing error caused by building sway for point-to-point FSO systems without IRS \cite{Alouini_Pointing, Steve_pointing_error}. Exploiting \eqref{Eq:hg_3D} and \eqref{Eq:Misalignment3D}, the PDF of $h_g$ can be obtained as
\begin{IEEEeqnarray}{lll} \label{Eq:PDF_h3D}
	f_{h_g}(h_g) =  &\frac{\varpi }{A_0}
	\left(\frac{h_g}{A_0}\right)^{\frac{(1+q^2)\varpi}{2q}-1}\times\nonumber\\
	  &I_0\left(-\frac{(1-q^2)\varpi}{2q}\ln\left(\frac{h_g}{A_0}\right)\right),
	 \quad 0< h_g \leq A_0,\quad 
	\quad
\end{IEEEeqnarray}
where $\varpi = \frac{(1+q^2)tw^2(d_{e2e})}{4q\Omega}$ is a constant and $\ln(\cdot)$ denotes the natural logarithm.

\section{Simulation Results}

\begin{figure}[!tbp]
	\centering
		\resizebox{0.9\linewidth}{!}{\psfragfig{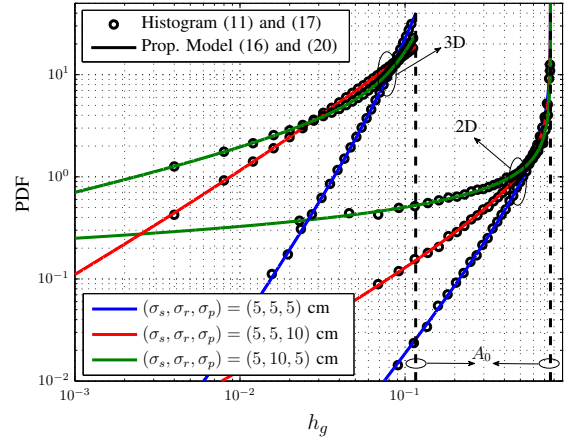}}   \vspace{-5mm}
		\caption{PDF of the GML for 2D and 3D scenarios.} \vspace{-4mm}
		\label{Fig:PDF}
\end{figure}

Unless stated otherwise, the default values of the parameter values used for 2D simulation are $\theta_b=\frac{\pi}{4}$, $\theta_{rs}=\frac{\pi}{10}$, $\theta_p=\frac{\pi}{3}$, $a_p=10$~cm, $a_r=50$~cm, $(x_r,y_r)=(400,400)$~m, and $(x_p,y_p)=(700,350)$~m. For 3D simulation, we use parameter values that are in-line with those for 2D, i.e., $\psi_r=\frac{\pi}{4}-\frac{\pi}{10}$, $\psi_p=\frac{\pi}{3}$, $d_{sr}=400\sqrt{2}$ m, $d_{rp}=50\sqrt{37}$~m, $a_p=10$~cm, and $a_r=50$~cm. Moreover, the simulation results reported in Fig.~\ref{Fig:PDF} were obtained based on Monte Carlo simulation and $10^6$ realizations of RVs $\epsilon_i^{j},i\in\{s,r,p\},j\in\{o,o_1,o_2\}$.

\begin{figure}[!tbp]
		\centering
		\resizebox{0.9\linewidth}{!}{\psfragfig{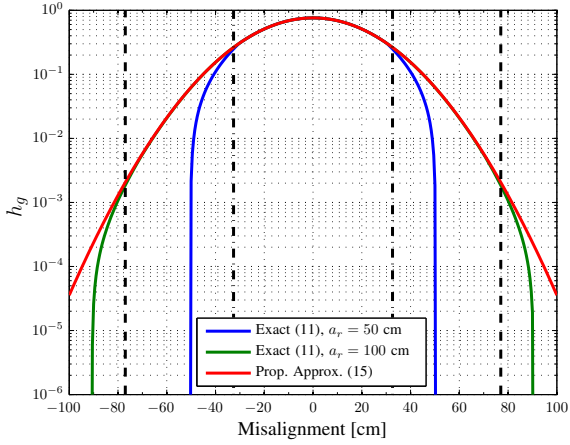}}  \vspace{-5mm}
		\caption{Conditional GML vs. misalignment $u$ for different sizes of IRS.} \vspace{-4mm}
		\label{Fig:CondGML}
\end{figure}

First, in Fig.~\ref{Fig:CondGML}, we study the impact of the size of the IRS on the conditional GML in \eqref{Eq:Pt}. In this figure, we show $h_g$ vs. misalignment $u$ for $a_r=50,100$~cm. As expected, we observe from Fig.~\ref{Fig:CondGML} that by increasing the misalignment magnitude ($|u|$), the channel gain $h_p$ decreases. Beam truncation occurs if the misalignment exceeds a cetrain critical value, i.e., when part of the PD is outside region $\mathcal{R}$, cf. \eqref{Eq:Truncated Gauss}.
In Fig.~\ref{Fig:CondGML}, we use dot-dashed (dashed) lines to denote this critical misalignment for $a_r=50$~cm ($a_r=100$~cm).
Fig.~\ref{Fig:CondGML} shows that the proposed approximation in \eqref{Eq:A0Approx} is accurate when beam truncation does not occur. However, since the approximation neglects beam truncation, it overestimates $h_g$ when beam truncation does occur. Moreover, we observe that, for $a_p=100$~cm, the impact of beam truncation manifests itself at larger values of $|u|$ compared to $a_p=50$~cm. Furthermore, Fig.~\ref{Fig:CondGML} shows that for a reasonable size of the IRS, i.e., $a_r>50$~cm, the proposed approximation is accurate even for large misalignment magnitudes, e.g. $|u|>35$~cm. Finally, we note that the PD receives no optical power, i.e., $h_g=0$, when none of the points on the PD surface belongs to $\mathcal{R}$, cf. Lemma~\ref{Lem:Truncated Gauss}.

Next, we study the accuracy of the proposed statistical models for 2D and 3D systems in \eqref{Eq:PDFs} and \eqref{Eq:PDF_h3D}, respectively. For the simulation results, we plot the histogram of $h_g$ given by \eqref{Eq:Pt} and \eqref{Eq:hg_3D} for 2D and 3D systems, respectively. Fig.~\ref{Fig:PDF} shows the PDF of $h_g$ for three fluctuation scenarios, namely Scenario~1: $(\sigma_s,\sigma_p,\sigma_r)=(5,5,5)$~cm where the building sways for the LS, IRS, and PD are similar; Scenario~2: $(\sigma_s,\sigma_p,\sigma_r)=(5,5,10)$~cm where the building sway for PD is larger than that for the LS and IRS\footnote{Scenario~2 yields the same results as scenario $(\sigma_s,\sigma_p,\sigma_r)=(10,5,5)$~cm due to the symmetry of the problem, see \eqref{Eq:Misalignment} and \eqref{Eq:Misalignment3D}.}; Scenario~3: $(\sigma_s,\sigma_p,\sigma_r)=(5,10,5)$~cm where the building sway for the IRS is larger than that for the LS and PD. Fig.~\ref{Fig:PDF} shows an excellent agreement between the proposed analytical statistical models and the simulation results. This is due to the fact that the impact of beam truncation is negligible as it occurs with small probability for the adopted system parameters. Moreover, we can observe from Fig.~\ref{Fig:PDF} that the building sway for the IRS has a larger impact than that for the PD (and LS). This is due to the factor $2\cos(\psi_{r})=1.782$ in \eqref{Eq:Misalignment} and \eqref{Eq:Misalignment3D} which enhances the variance of the corresponding building sway. 

\section{Conclusions}

In this paper, we proposed IRS-based FSO systems in order to relax the LOS requirement of conventional FSO systems. We developed corresponding conditional and statistical channel models which characterize the impact of the physical parameters of the IRS, such as its size, position, and orientation, on the quality of the end-to-end FSO channel. These channel models can be used for performance analysis of IRS-based FSO systems. Simulation results confirmed the validity of the developed channel models for typical IRS sizes (i.e., $a_r>50$~cm) where beam truncation is negligible. Furthermore, our results showed that depending on the angle between the beam direction and the IRS plane, building sway for the IRS could have a larger impact on the quality of the end-to-end FSO channel than building sway for the Tx and Rx for angles smaller than $\pi/3$.

\iftoggle{ConfVersion}{%
}{%

\appendices

\section{}\label{App:Lem_Truncated Gauss}
The ray that originates at the virtual LS, i.e., at point $\mathbf{p}_{\mathrm{vs}}$, and intersects with the upper corner of the IRS, i.e., $(x_r+\cos(\theta_{rs})a_r,y_r+\sin(\theta_{rs})a_r)$, is given by $y=\frac{y_r+\sin(\theta_{rs})a_r-y_{vs}}{x_r+\cos(\theta_{rs})a_r-x_{vs}}(x-x_{vs})+y_{vs}$. Similarly, the ray that originates at $\mathbf{p}_{\mathrm{vs}}$ and intersects the lower corner of the IRS, i.e., $(x_r-\cos(\theta_{rs})a_r,y_r-\sin(\theta_{rs})a_r)$, is given by $y=\frac{y_r-\sin(\theta_{rs})a_r-y_{vs}}{x_r-\cos(\theta_{rs})a_r-x_{vs}}(x-x_{vs})+y_{vs}$. Within these two lines the power density is non-zero and the corresponding region is defined by $\mathcal{R}$. Now, we find the beam footprint center on line $L_1$ which is perpendicular to the beam line and passes through point $\tilde{\mathbf{p}}=(\tilde{x},\tilde{y})$, i.e., $y=\cot(\theta_b-2\theta_{rs})(x-\tilde{x})+\tilde{y}$. To do so, we calculate the intersection of the beam line with line $L_1$ denoted by $\mathbf{p}_{\mathrm{c}}$. Next, we calculate the distances $r$ and $d$. In particular, the distance between $\mathbf{p}_{\mathrm{c}}$ and $\tilde{\mathbf{p}}$ determines $r$ and the distance between $\mathbf{p}_{\mathrm{vs}}$ and $\mathbf{p}_{\mathrm{c}}$ yields $d$, cf. Lemma~\ref{Lem:Truncated Gauss}. This completes the proof.

\section{}\label{App:Prop_TotalFrac}

The total fraction of power collected by the PD can be obtained by integrating over the power density on the PD line. For ease of notation, we define variable $\rho$ as the distance between any point $\mathbf{p}$ on the PD line and the beam footprint center. In addition, let $I_{\mathrm{rfl}}(\rho)\mathrm{d}\rho$ denote the fraction of power collected on the infinitesimally small line $\mathrm{d}\rho$, i.e., $\mathrm{d}\rho\to 0$, on the PD. Next, we relate $I_{\mathrm{rfl}}(\rho)$ to the power density expression given in \eqref{Eq:Truncated Gauss}. In particular, \eqref{Eq:Truncated Gauss} is a function of two variables, $r$ and $d$, denoted by $I^{\mathrm{orth}}_{\mathrm{rfl}}(r;d)$. We can obtain $r$ as $\sin(\psi)\rho$ and bound $d$ as follows
\begin{IEEEeqnarray}{lll} \label{Eq:d_Bound}
\sqrt{d_{e2e}^2-L_c^2}-a_p\leq d\leq \sqrt{d_{e2e}^2-L_c^2}+a_p.
\end{IEEEeqnarray}
Assuming $a_p,L_c\ll d_{e2e}$, we can safely approximate $d$ as $d\approx d_{e2e}$. Therefore, we obtain  $I_{\mathrm{rfl}}(\rho)\mathrm{d}\rho=\sin(\psi)I^{\mathrm{orth}}_{\mathrm{rfl}}(\sin(\psi)\rho;d_{e2e})\mathrm{d}\rho$, where $\psi=\theta_b+\theta_p-2\theta_{rs}$ is the angle between the beam direction and the PD line and the term $\sin(\psi)$ is due to the non-orthogonality of the PD.
 The total fraction of power captured by the PD is obtained by integrating $I_{\mathrm{rfl}}(\rho)$ over the PD line as
\begin{IEEEeqnarray}{lll} \label{Eq:Pt_App}
h_g=\begin{cases}\displaystyle 
\int_0^{\rho_1}I_{\mathrm{rfl}}(\rho)\mathrm{d}\rho+\int_0^{\rho_2}I_{\mathrm{rfl}}(\rho)\mathrm{d}\rho,& \mathrm{if}\,\,\rho_{12}=2a_p\\
\displaystyle 
\Big|\int_0^{\rho_1}I_{\mathrm{rfl}}(\rho)\mathrm{d}\rho-\int_0^{\rho_2}I_{\mathrm{rfl}}(\rho)\mathrm{d}\rho\Big|,& \mathrm{otherwise,}
\end{cases}\quad
\end{IEEEeqnarray}
where $\rho_1$, $\rho_2$, and $\rho_{12}$ are given in Proposition~\ref{Prop:TotalFrac}. The two cases in \eqref{Eq:Pt_App} correspond to whether or not the center of the beam footprint lies on the PD. The integrals in \eqref{Eq:Pt_App} can be computed as 
\begin{IEEEeqnarray}{lll}\label{Eq:Int} 
\int_0^{\rho_i}I_{\mathrm{rfl}}(\rho)\mathrm{d}\rho=\int_0^{\rho_i}\frac{2\sin(\psi)}{\pi w^2(d_{e2e})}\exp\left(-\frac{2(\sin(\psi)\rho)^2}{w^2(d_{e2e})}\right)\mathrm{d}\rho\nonumber\\
=\frac{1}{\sqrt{2\pi}w(d_{e2e})}\mathrm{erf}\left(\frac{\sqrt{2}\sin(\psi)\rho_i}{w(d_{e2e})}\right).
\end{IEEEeqnarray}
Substituting \eqref{Eq:Int} into \eqref{Eq:Pt_App} leads to \eqref{Eq:Pt} and concludes the proof.

\section{}\label{App:A0Approx}
From Appendix~\ref{App:Prop_TotalFrac}, we have $I_{\mathrm{rfl}}(\rho)\mathrm{d}\rho=\sin(\psi)I^{\mathrm{orth}}_{\mathrm{rfl}}$ $(\sin(\psi)\rho;L)\mathrm{d}\rho$. Moreover, the distance between the PD center and the beam footprint center is denoted by $u$. The fraction of power that is collected by the PD is obtained by integrating $I_{\mathrm{rfl}}(\rho)$ over the PD line as follows 
\begin{IEEEeqnarray}{lll} \label{Eq:A0New}
h_g&=\int_{-a_p-u}^{a_p-u}\sin(\psi)I^{\mathrm{orth}}_{\mathrm{rfl}}(\sin(\psi)\rho;L)\mathrm{d}\rho\nonumber\\
&\overset{(a)}{=}\int_{-a_p-u}^{a_p-u}
\frac{2\sin(\psi)}{\pi w^2(L)} \exp\left(-\frac{2\sin^2(\psi)\rho ^2}{w^2(L)}\right)\mathrm{d}\rho,
\end{IEEEeqnarray}
where equality $(a)$ is the result of replacing $I^{\mathrm{orth}}_{\mathrm{rfl}}(\cdot)$ with \eqref{Eq:Truncated Gauss}. In order to approximate $h_g$, we first use the Taylor series expansion of the exponential term as follows
\begin{IEEEeqnarray}{lll} \label{Eq:delta}
h_g&= \frac{2\sin(\psi)}{\pi w^2(L)}\sum_{n=0}^{\infty}\int_{-a_p-u}^{a_p-u}
\frac{\left(-\frac{2\sin^2(\psi)\rho ^2}{w^2(L)}\right)^n}{n!}\mathrm{d}\rho\nonumber\\
&=\sum_{n=0}^{\infty}\frac{\frac{2\sin(\psi)}{\pi w^2(L)}\left(-\frac{2\sin^2(\psi)}{w^2(L)}\right)^n}{n!(2n+1)}\left((a_p-u)^{2n+1}+(a_p+u)^{2n+1}\right)\nonumber\\
&=\sum_{k=0}^{\infty}A_{2k}\left(\frac{\sqrt{2}\sin(\psi)u}{w(d_{e2e})}\right)^{2k},
\end{IEEEeqnarray}
where $A_{2k}=\sum_{n=k}^{\infty}\frac{2\sqrt{2}(-1)^{n}{2n+1\choose 2k}}{\pi w(d_{e2e})n!(2n+1)}\left(\frac{\sqrt{2}\sin(\psi)a_p}{w(d_e2e)}\right)^{2n+1-2k}$. By equating the first two terms of \eqref{Eq:delta} to the same terms in the Taylor series expansion of a Gaussian pulse of form $c\exp\left(\frac{-2u^2}{tw^2(d_{e2e})}\right)$, we obtain \eqref{Eq:A0Approx}. This completes the proof.
}


\bibliographystyle{IEEEtran}
\bibliography{My_Citation_01-05-2019}

\end{document}